\documentclass[showpacs,amsmath,amssymb,aps]{revtex4}
\usepackage{graphicx}
\usepackage{dcolumn}
\usepackage{epsfig}
\usepackage{bm}
\usepackage{gensymb}
\usepackage{textcomp} 
\usepackage{color}

\begin{document}

\title{Quasielastic electron- and neutrino-nucleus scattering in a continuum random phase approximation approach}

\author{V.~Pandey\footnote{Vishvas.Pandey@UGent.be}, N.~Jachowicz, T.~Van Cuyck, J.~Ryckebusch, M.~Martini}

\affiliation{Department of Physics and Astronomy,\\ Ghent University, \\Proeftuinstraat 86, \\ B-9000 Gent, Belgium.\\}

\begin{abstract}

We present a continuum random phase approximation approach to study electron- and neutrino-nucleus scattering cross sections, 
in the kinematic region where quasielastic scattering is the dominant process. We show the validity of the formalism by 
confronting inclusive ($e,e'$) cross sections with the available data. We calculate flux-folded cross sections for charged-current 
quasielastic antineutrino scattering off $^{12}$C and compare them with the MiniBooNE cross-section measurements. We pay special 
emphasis to the contribution of low-energy nuclear excitations in the signal of accelerator-based neutrino-oscillation experiments.
 
\end{abstract}


\maketitle

\section{Introduction}

Recent years have seen an enormous enhancement in the understanding of neutrino-oscillation parameters in accelerator-based experiments.
These experiments are however confronted with a number of problems. These are related to the large systematic uncertainties associated 
with the neutrino-nucleus signal in the detector. Major issues arise from the fact that the neutrino energy-flux in experiments is 
distributed over a wide range of energies from very low to a few GeV. Hence a number of nuclear effects over a broad kinematical range 
(from low-energy nuclear excitations to multinucleon emission) simultaneously come into play. The simulation codes used in the analysis 
of the experimental results are predominantly based on relativistic Fermi gas (RFG) models. RFG can describe the quasielastic (QE) cross 
section sufficiently accurate for medium momentum ($q \approx $ 500 MeV/c) transfer reactions, but its description becomes poor for low 
momentum ($q \lesssim $ 300 MeV/c) transfer processes, where nuclear effects are prominent. For the broad neutrino energy-flux used in the 
experiments, more realistic models are required. 

In this work, we present a self-consistent continuum random phase approximation (CRPA) approach to calculate QE electron and 
neutrino-scattering cross-sections off the nucleus. This formalism was used to describe exclusive photo-induced and electron-induced QE 
scattering~\cite{Jan:1988, Jan:1989}, inclusive neutrino scattering at supernova 
energies~\cite{Natalie:nc1999, Natalie:cc2002, Natalie:nc2004, Natalie:sn2006, Natalie:proc2009, Natalie:nuintproc2012} and charged-current 
quasielastic (CCQE) antineutrino scattering at intermediate energies~\cite{Vishvas:ccqeantinu2014}. We will briefly describe the essence of 
our model, for an updated version of the formalism we refer the reader to Ref.~\cite{Vishvas:crpa2014}. The main update in 
Ref.~\cite{Vishvas:crpa2014} from Ref.~\cite{Vishvas:ccqeantinu2014}, are the inclusion of relativistic corrections and a suppression of the 
RPA quenching at high $Q^2$. We start with a mean-field (MF) description of the nucleus where we solve the Hartree-Fock (HF) equations with 
a Skyrme (SkE2) two-body interaction~\cite{Jan:1989, Waroquier:1987} to obtain the MF potential. We obtain the continuum wave functions by 
integrating the positive energy Schr\"odinger equation with appropriate boundary conditions, hence taking into account final-state interactions 
in this manner. Long-range correlations are implemented by means of a CRPA approach based on a Green's function formalism. The polarization
propagator is approximated by iteration of its first-order contribution. In this way, the formalism takes into account one-particle
one-hole excitations out of the correlated nuclear ground state. Within the RPA an excited nuclear state is represented as the coherent 
superposition of the particle-hole ($ph^{-1}$) and hole-particle ($hp^{-1}$) excitations out of a correlated ground state
\begin{equation}
 \arrowvert \Psi_{RPA}^{C} \rangle = \sum_{C'} \left[ X_{C, C^{'}} ~ \arrowvert p'h'^{-1} \rangle - 
 ~Y_{C, C^{'}}~ \arrowvert h'p'^{-1} \rangle \right]~,
\end{equation}
where $C$ denotes all quantum numbers identifying an accessible channel. The RPA polarization propagator can be written as
\begin{eqnarray}
 & & \Pi^{(RPA)} (x_1,x_2;E_x) =  \Pi^{(0)} (x_1,x_2;E_x)  + \frac{1}{\hbar} \int dx dx' \Pi^{0} (x_1,x;E_x)  
\tilde{V}(x, x') \Pi^{(RPA)} (x',x_2;E_x), \nonumber \\ && \label{propagator}
\end{eqnarray}
where $E_x$ is the excitation energy of the nucleus and $x$ is a short-hand notation for the combination of spatial, spin and isospin 
coordinates. The $\Pi^{(0)}$ corresponds to the MF contribution and $\tilde{V}$ is the antisymmetrized nucleon-nucleon interaction.

We used the modified effective momentum approximation (MEMA)~\cite{Engel:1998}, in order to take into account the influence of the
nuclear Coulomb field on the ejected lepton. In order to prevent the SkE2 force from becoming unrealistically strong at high virtuality 
$Q^{2}$, we introduce a dipole hadronic form factor at the nucleon-nucleon interaction vertices~\cite{Vishvas:crpa2014}. Further, we have 
implemented relativistic kinematic corrections~\cite{Donnelly:1998} in an effective manner. 

We first test the reliability of the formalism by confronting ($e,e'$) scattering cross sections with the data of 
Refs.~\cite{edata12C:Barreau, edata12C:Sealock, edata12C:Bagdasaryan, edata12C:Day}. Thereby, we present updated results 
of flux-folded charged-current quasielastic (CCQE) antineutrino scattering off $^{12}$C and compare them with the MiniBooNE 
measurements~\cite{miniboone:ccqeantinu}. Further, we discuss the contribution of neutrino-induced low-energy nuclear excitations 
in the signal of the accelerator-based neutrino oscillation experiments.

\begin{figure}
\center
\includegraphics[width=0.99\columnwidth]{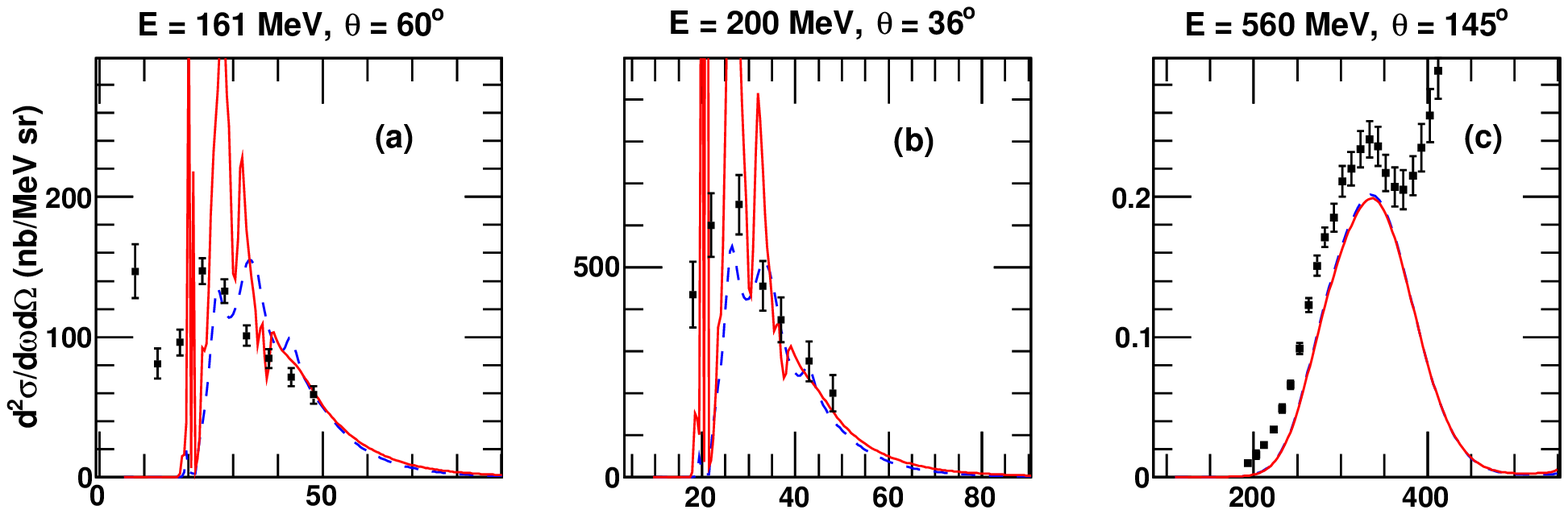}
\includegraphics[width=0.99\columnwidth]{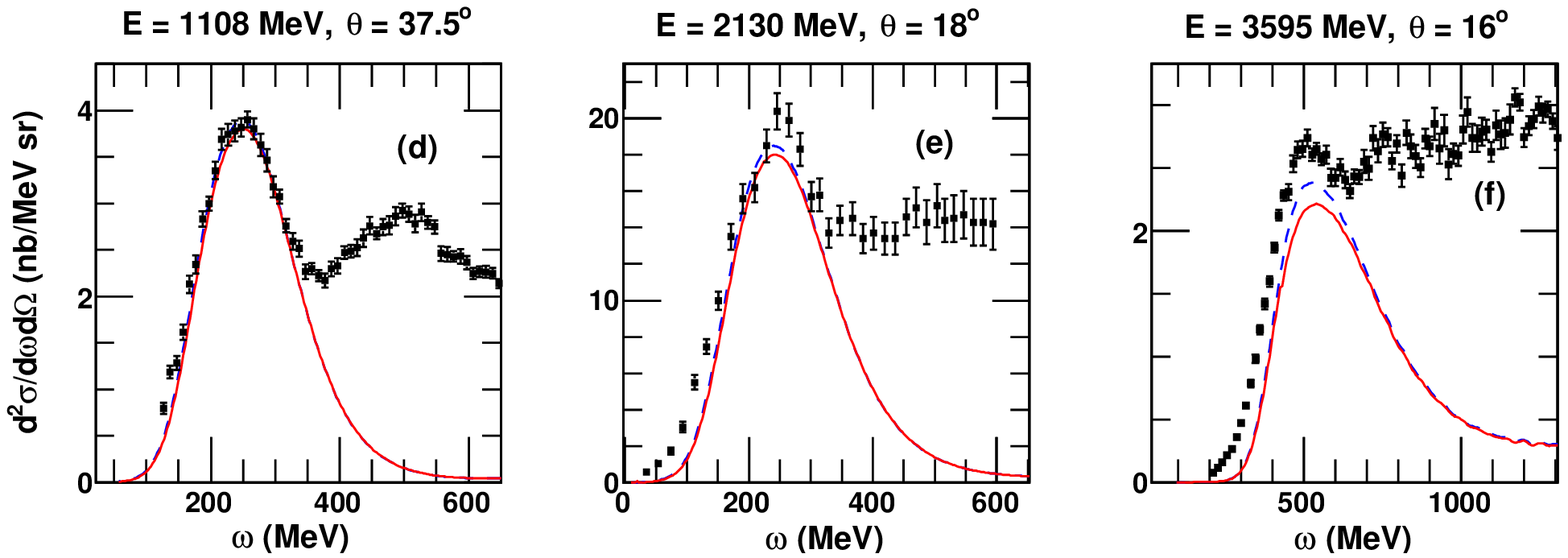}
\caption{The double-differential cross-sections for $^{12}$C($e,e'$) plotted as a function of excitation energy $\omega$. 
The incident electron energy $E$ and lepton scattering angle $\theta$ are listed on top of each panel.
Solid lines are CRPA cross sections and dashed-lines are HF cross sections. Experimental data is taken from 
Refs.~\cite{edata12C:Barreau, edata12C:Sealock, edata12C:Bagdasaryan, edata12C:Day}.}
\label{fig_e_scattering}
\end{figure}

\section{Cross section results}

We start this section by showing some examples of electron-scattering results.
In Fig.~\ref{fig_e_scattering}, we show our prediction of QE $^{12}$C($e,e'$) scattering cross-sections and compare them with the 
measurements of the Refs.~\cite{edata12C:Barreau, edata12C:Sealock, edata12C:Bagdasaryan, edata12C:Day}. Our predictions successfully
describe the data over the broad kinematical range considered here. The formalism successfully describes low-energy excitations 
(panel (a) and (b)) below the QE peak. The forward scattering cross sections, even for higher incoming electron energies, are dominated by 
the QE contribution. However, the data include cross section contributions beyond the QE channel, like $\Delta$-excitations and other inelastic 
channels. Our calculations are intended to predict only the QE behavior. A detailed comparison of ($e,e'$) cross section on $^{12}$C,  $^{16}$O
and $^{40}$Ca is performed in Ref.~\cite{Vishvas:crpa2014}. An overall successful description of QE ($e,e'$) cross section data and especially
low-energy excitations, validates the reliability of our formalism.

\begin{figure}
\center
\includegraphics[width=0.99\columnwidth]{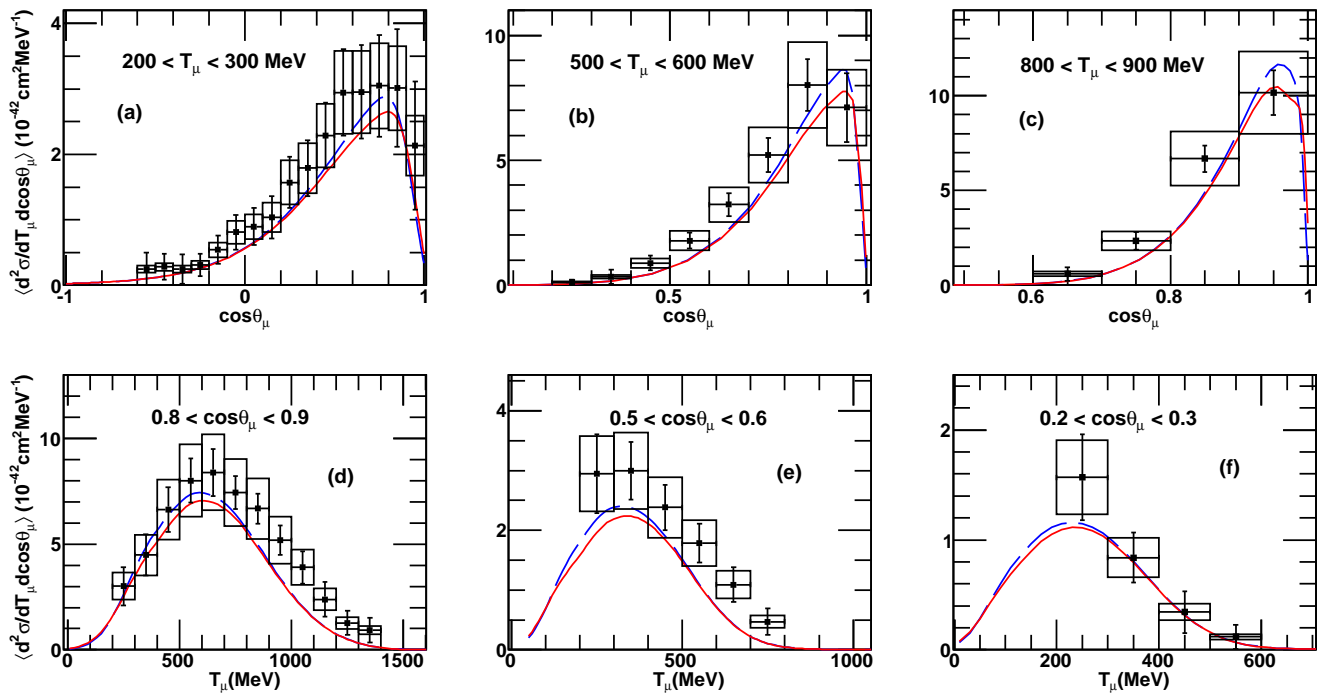}
\caption{(Color online) MiniBooNE flux-folded double-differential cross section per target proton 
for $^{12}$C$(\bar{\nu}_{\mu},  \mu^{+})X$ plotted as a function of the direction of the scattering
lepton $\cos\theta_{\mu}$ for different values of its kinetic energy
$T_{\mu}$ values (top),  as a function $T_{\mu}$ for different ranges of $\cos\theta_{\mu}$ (bottom).  
Solid lines are CRPA and dashed lines are HF calculations. MiniBooNE data~\cite{miniboone:ccqeantinu}
are filled squares, error bars represent the shape uncertainties and error boxes represent the 17.2\% 
normalization uncertainty.}
\label{fig_MiniBooNE_anu_ddxs}
\end{figure}

We show the double-differential cross section for $^{12}$C$(\bar{\nu}_{\mu},  \mu^{+})X$, folded with the MiniBooNE antineutrino 
flux~\cite{miniboone:ccqeantinu}, in Fig.~\ref{fig_MiniBooNE_anu_ddxs}. The top panels show the cross section in $T_{\mu}$ bins and the
bottom panels show the cross section in $\cos\theta_{\mu}$ bins. The cross section is integrated over the corresponding bin width.
We adopt an axial mass value of $M_A$ = 1.03 GeV, in the dipole axial form factor. HF and CRPA cross sections are compared with the 
MiniBooNE measurements of Ref.~\cite{miniboone:ccqeantinu}. MiniBooNE data is presented with both shape and normalization uncertainties. 
Overall, CRPA and HF calculations successfully reproduce the gross features of the measured cross section. The predictions tend to underestimate 
the data. It has been suggested in Refs.~\cite{Martini:2009, Martini:2010, Nieves:2011, Martini:2013, Nieves:2013} that the inclusion of 
multinucleon contributions, which are not included in our calculations, are essential for a more complete reproduction of the data.

\begin{figure}
\center
\includegraphics[width=0.67\columnwidth]{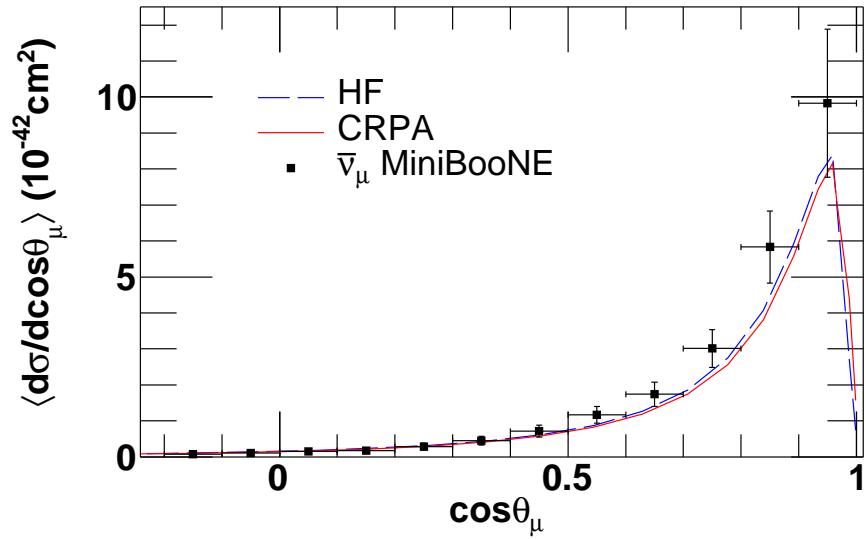}
\caption{MiniBooNE flux-folded cross section per target proton for $^{12}$C$(\bar{\nu}_{\mu},
\mu^{+})X$ as a function of $\cos\theta_{\mu}$. The MiniBooNE data~\cite{miniboone:ccqeantinu} are 
integrated over $T_{\mu}$.}
\label{fig_MiniBooNE_anu_dxs} 
\end{figure}

\begin{figure}
\center
\includegraphics[width=0.67\columnwidth]{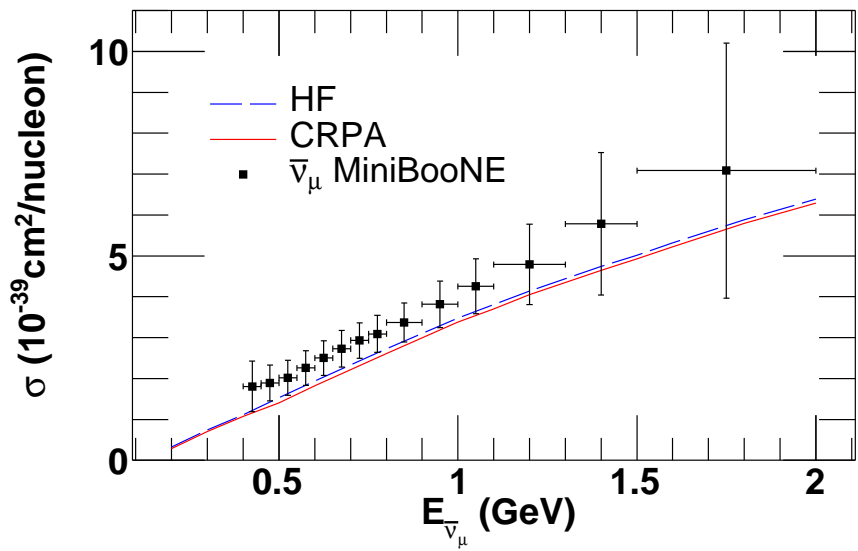}
\caption{Flux unfolded total cross section per target proton for $^{12}$C$(\bar{\nu}_{\mu},
\mu^{+})X$ as a function of $E_{\bar{\nu}_{\mu}}$, compared with MiniBooNE data~\cite{miniboone:ccqeantinu}.}
\label{fig_MiniBooNE_anu_xs}
\end{figure}

The flux-folded differential cross section as a function of $\cos\theta_{\mu}$, is shown in Fig.~\ref{fig_MiniBooNE_anu_dxs}.
For comparison with data, we integrate MiniBooNE data over $T_{\mu}$. It is interesting to note that in the very forward direction 
the CRPA results are larger than the HF ones. This is due to the collective giant resonance contributions which are absent in the HF 
approximation but appear in CRPA results that include long-range correlations. In Fig.~\ref{fig_MiniBooNE_anu_xs}, we present
total cross sections per target proton as a function of neutrino energy and compare them with the experimental data. Unlike double-differential
cross sections, this quantity is model dependent. The theoretical calculations are function of a true antineutrino energy while the experimental 
data are function of reconstructed antineutrino energy. Up to $E_{\bar{\nu}}$ = 0.4 GeV, the HF results essentially coincide with the CRPA ones. 
This is due to a compensation between a reduction in the QE region and an enhancement in the giant resonance part of the CRPA results.
For $E_{\bar{\nu}} \gtrsim$ 0.4 GeV, the CRPA results are slightly smaller than the HF ones. 

\begin{figure}
\center
\includegraphics[width=0.9\columnwidth]{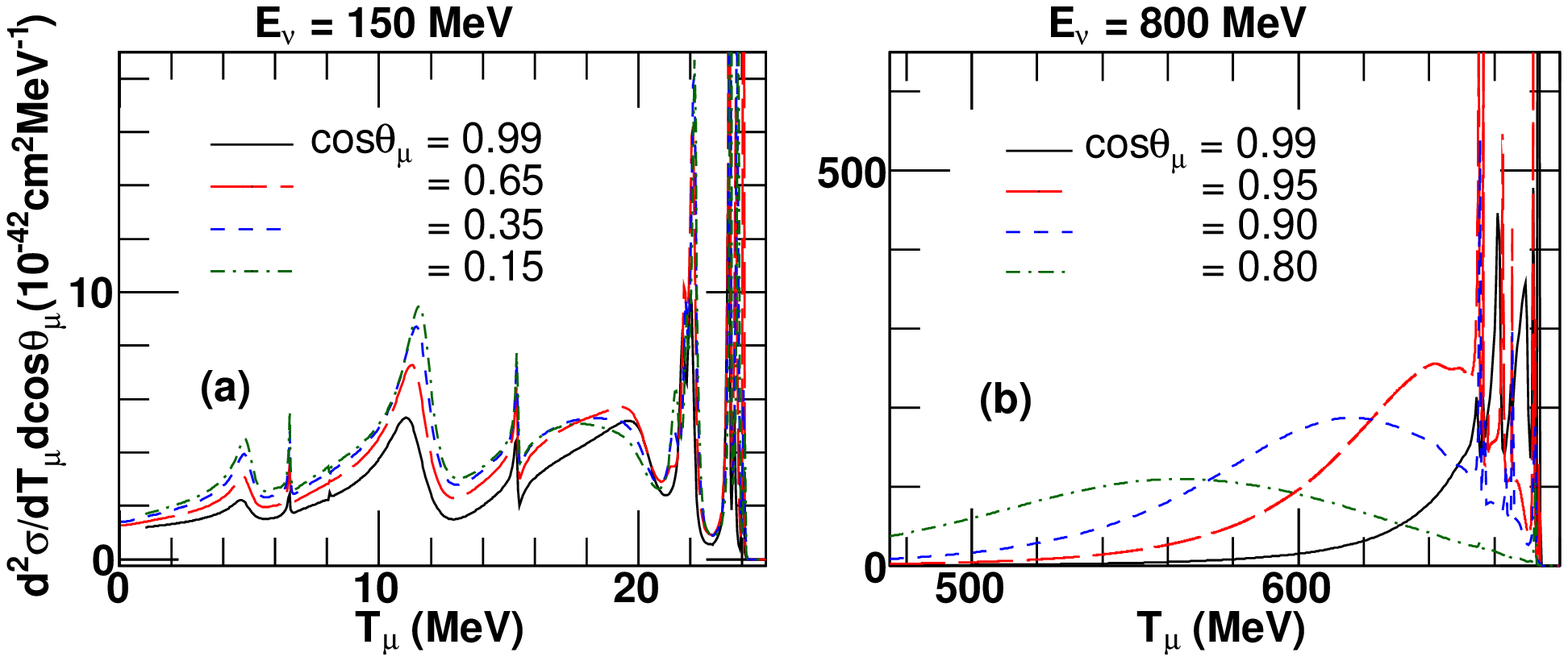}
\caption{Neutrino induced low-energy nuclear excitations in double differential cross section for
$^{12}$C($\nu_{\mu}, \mu^{-}$) plotted as a function $T_{\mu}$, for different $\cos\theta_{\mu}$ 
values.}
\label{fig_lowenergy}
\end{figure}

In order to illustrate the impact of the low-energy nuclear excitations, in Fig.~\ref{fig_lowenergy}, we show the double-differential 
cross-section for fixed neutrino energies and fixed scattering angles. As it appears in panel (a), 150 MeV energy neutrinos
induce low-lying nuclear excitations at all scattering angles. For neutrino energies of 800 MeV, which is near the mean energy of the
MiniBooNE~\cite{miniboone:ccqenu} and T2K~\cite{t2k:qenu} fluxes, in panel (b), the forward scatterings still show sizable low-energy 
excitations. This feature can have a non-negligible contribution to the neutrino signals in these experiments, but can not be accounted 
for within the RFG-based simulation codes. As already mentioned, one can observe in Fig.~\ref{fig_MiniBooNE_anu_dxs}, that at very forward 
scatterings, $\cos\theta_{\mu} \approx$ 1, the CRPA cross section generates more strength (emerging from the low-lying excitation) than the HF.

\section{Conclusions}

We have presented a continuum random phase approximation approach for quasielastic electron- and neutrino-nucleus scattering. We validated the
reliability of our formalism, in the quasielastic region, by comparing ($e,e'$) cross section with the available data. An interesting feature of
our CRPA formalism is the successful prediction of low-energy nuclear excitations. We calculated flux-folded $^{12}$C$(\bar{\nu}_{\mu},  \mu^{+})X$ 
cross sections and compared them with the MiniBooNE antineutrino cross-section measurements. CRPA predictions are successful in describing the gross 
features of the cross section but seem to underestimate slightly the measured cross section. We illustrated how low-energy nuclear excitations can 
possibly account for non-negligible contributions to the neutrino signal in accelerator-based neutrino-oscillation experiments.

\acknowledgments
This research was funded by the Interuniversity Attraction Poles Programme initiated by the Belgian 
Science Policy Office, the Research Foundation Flanders (FWO-Flanders) and by the Erasmus Mundus External Cooperations Window's Eurindia Project.

\end{document}